\documentclass[a4paper,11pt]{article}
\usepackage{pos}
\usepackage{bm}
\usepackage{graphicx}

\title{Next challenges in semi-leptonic B decays}

\author*[a,b]{Shoji Hashimoto}

\affiliation[a]{Institute for Particle and Nuclear Studies,
  High Energy Accelerator Research Organization (KEK),
  Tsukuba, Ibaraki 305-0801, Japan}

\affiliation[b]{SOKENDAI (The Graduate University for Advanced Studies),
  Tsukuba, Ibaraki 305-0801, Japan}

\emailAdd{shoji.hashimoto@kek.jp}

\abstract{
  Lattice calculation of $B\to D^{(*)}\ell\bar\nu$ decay form factors is
  now available from multiple lattice collaborations for both
  zero-recoil limit and non-zero recoil kinematics.
  Yet, there are a number of challenges we face to better control the
  systematic errors and to extend the application of lattice
  calculations to related quantities.
  The subjects are chosen from my own perspective and not meant to be
  conprehensive. 
}

\FullConference{European network for Particle physics, Lattice field theory and Extreme computing (EuroPLEx2023)\\
 11-15 September 2023\\
Berlin, Germany\\}

\begin{document}
\maketitle

\section{$|V_{cb}|$ and $|V_{ub}|$: current position in the history}
The CKM matrix element $|V_{cb}|$ is determined through the
semi-leptonic decays of the $B$ meson.
There are two kinematical variables (apart from those giving angular
distributions), {\it i.e.} the invariant mass squared $m_X^2$ and the
recoil momentum squared $\bm{q}^2$ of the charmed hadronic
state (potentially multi-hadron state) in the final state.
The $|V_{cb}|$ can be determined at any point of the
two-dimensional phase space or from integrated decay rate over the
phase space with an arbitrary weight, provided that reliable
theoretical calculation is available for the QCD effect,
from lattice QCD in particular. 
The discussion of the inclusive method, which uses the integrated
decay rate over the entire phase space, is left for a companion talk
\cite{Hashimoto_inclexcl}.

In the early days of the $|V_{cb}|$ determination from exclusive
processes $B\to D^{(*)}\ell\bar\nu$, one relied on the
heavy quark symmetry, that says that the relevant form factor is
normalized to unity at the zero-recoil point
\cite{Isgur:1989vq,Isgur:1990yhj}. 
The correction of order $1/m_c$ vanishes at this kinematical point,
and the correction starts only at order $1/m_c^2$ (Luke's theorem)
\cite{Luke:1990eg}. 
Lattice QCD can benefit from the same symmetry argument
\cite{Hashimoto:1999yp,Hashimoto:2001nb},
and the calculation is actually done  the remaining $1/m_c^2$
corrections
(see \cite{MILC:2015uhg,FermilabLattice:2014ysv}, for example).
The value of $|V_{cb}|$ thus obtained combined with the experimental
results extrapolated to the zero-recoil point
is in tension with the corresponding determination from the
inclusive decays,
{\it i.e.} the integral over the whole phase space of the
semi-leptonic decays.
In order to understand the source of the problem, it is 
important to extend the determination of the exclusive 
processes toward non-zero recoil kinematics, since the existing
$|V_{cb}|$ determination utilizes only a tiny corner of the phase
space. 
It is desirable to use the excited states to check with the broader
two-dimensional phase space, so that they finally cover the entire
phase space.

Recently, the lattice calculation of $B\to D^{(*)}\ell\bar\nu$ form
factors has been extended to the non-zero recoil
\cite{FermilabLattice:2021cdg,Harrison:2023dzh,Aoki:2023qpa},
and some confusion emerged.
The slope of the form factor $\mathcal{F}(w)$ near the zero recoil limit
$w=1$ seems significantly steeper in the Fermilab/MILC
\cite{FermilabLattice:2021cdg} and HPQCD \cite{Harrison:2023dzh}
calculations compared to the experimental data by Belle
\cite{Belle:2018ezy,Belle:2017rcc,Belle:2023bwv}.
Indeed, the overall shape of the form factor as a function of $w$ and
angular observables seems clearly inconsistent with the Belle data.
The calculation by JLQCD \cite{Aoki:2023qpa} is in
agreement with Belle, on the other hand.
Such inconsistency among different groups suggests some unknown
systematics in the current lattice calculations.

The $B\to\pi\ell\bar\nu$ form factors to be used in the determination of
$|V_{ub}|$ have also be computed on the lattice
\cite{FermilabLattice:2015mwy,Flynn:2015mha,Colquhoun:2022atw};
consistency among lattice groups is better albeit relatively large
errors.
Inconsistency between exclusive and inclusive determinations is still
to be understood also here.

In the following sections, I discuss what would be the next
challenges, 
in order to understand the discrepancy present in the $B\to D^{(*)}$
form factors, in particular.
I also mention some other interesting application of the lattice
calculation.

\section{Next challenge I: $1/m_Q$ corrections}
For the decays of heavy mesons, the heavy quark symmetry provides
useful constraints when analyzing the lattice results.
It dictates that the form factor remains constant in the heavy quark
limit after taking out an appropriate scaling factor of the form
$m_Q^\alpha$ with a known power $\alpha$.
Away from the infinite quark mass limit, the correction of order
$1/m_Q$ is expected for finite heavy quark mass $m_Q$
(unless some symmetry constraint exists, such as Luke's theorem).

In the lattice calculation, one has to deal with the
physical quark mass dependence of the form $1/m_Q$ together with
unphysical discretization effect, which may have the form $(am_Q)^n$
for a finite lattice spacing $a$.
Ideally, one should take the continuum limit before applying the fit
of the physical $1/m_Q$ dependence, but practically the analysis is
done with a combined fit of both $1/m_Q$ and $am_Q$ dependences.

\begin{figure}[tbp]
  \centering
  \includegraphics[width=8cm]{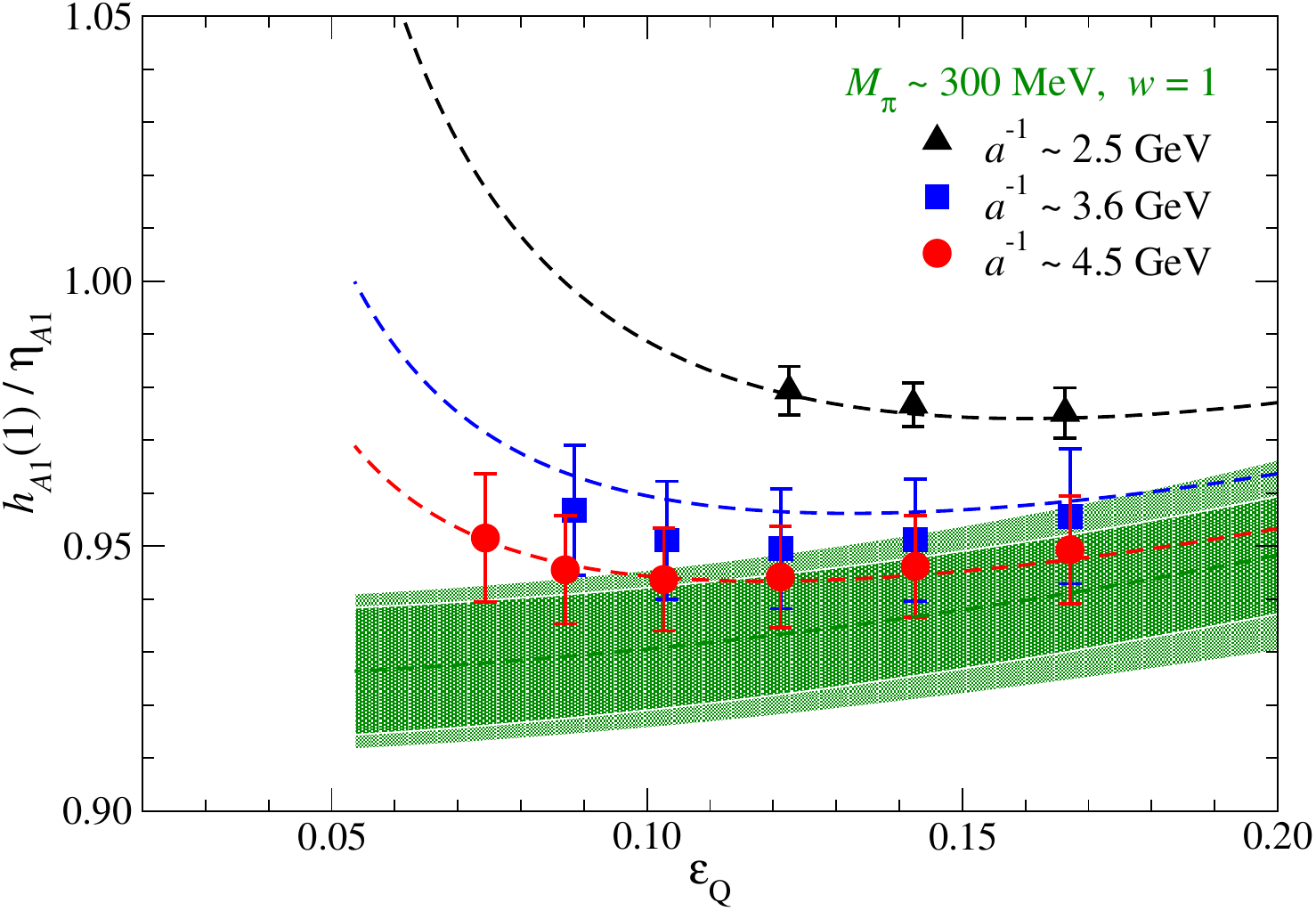}
  \caption{$B\to D^*$ form factor $h_{A1}(w)$ in the zero recoil limit
    $w=1$ plotted as a function of $\epsilon_Q=\bar\Lambda/2m_Q$ with
    $\bar\Lambda$ = 500~MeV.
    The plot is from JLQCD \cite{Aoki:2023qpa}, and the data at three
    lattice spacings are shown.
    Green band represents the continuum limit extrapolated from the
    lattice data.
  }
  \label{fig:BtoDstar}
\end{figure}

An example is shown for the $B\to D^*$ form factor $h_{A1}(w)$ in 
Fig.~\ref{fig:BtoDstar}, where the zero-recoil form factor is plotted
as a function of $\bar\Lambda/2m_Q$ with $\bar\Lambda$ = 500~MeV.
The lattice data at three lattice spacings show a significant
dependence on $a$, and the fit curve going through each $a$ shows a
projection of the global fit on that particular value of $a$.
It diverges towards $m_Q\to\infty$ because of the expected form
$(am_Q)^2$, while the continuum limit (the green band) is finite.
To control such a delicate extrapolation to the desired precision,
which is in this case better then 1\%, more data at finer lattice
spacings would be helpful.
(In this particular case, a fit for different sets of data with
limited value of $am_Q$ has been attempted to confirm the stability.)

The issue becomes even more delicate when we consider the dependence
on the momentum transfer in the analysis.
The momentum transfer $q^2$ is defined by the four-vector $q_\mu$ of
the virtual $W$ boson, and it takes a value between 0 and
$(m_B-m_{D*})^2$ depending on the kinematical configuration of the
decay.
The heavy-quark limit must be taken for appropriately rescaled form
factors and a proper kinematical variable.
For the $B\to D^{*}$ form factors, they are the HQET-motivated form
factors $h_i(w)$ as a function of $w=v\cdot v'$, an inner product of
initial and final heavy meson velocities.
The heavy quark extrapolation must be done for a fixed value of $w$.
The phase space increases as $m_Q$, and the region of larger $w$ (or
the region near $q^2=0$) is not accessible with small $m_Q$.

It is common to use the $z$-parametrization of the form factors
\cite{Bourrely:2008za} to represent the overall shape of the form
factors instead of parametrizing the $q^2$ dependence by its Taylor
expansion.
The $z$-parametrization relies on the analyticity of the form factor
and some unitarity bounds.
It is believed that only a few leading-order terms in a
$z$-expansion is enough to describe the entire $q^2$ range relevant
for the semi-leptonic decays.
Unfortunately, the analyticity does not tell anything about the
dependence on $m_Q$, 
and the heavy-quark extrapolation assuming the $1/m_Q$ dependence
must be performed before applying the $z$-expansion at the physical $b$
quark mass.
(The analysis of \cite{Aoki:2023qpa} is done in that way.)
The combined fit with a double expansion in terms of
$z$ and $1/m_Q$ as employed, {\it e.g.}, in \cite{Harrison:2021tol}
is not consistent with the heavy-quark scaling, rigorously speaking.
Although the ansatz might still give a reasonably good description of
the lattice data in the region of available data, the associated error
when extrapolated to the region of $1/m_Q$ that the lattice data do
not cover should be carefully examined.

\section{Next challenge II: excited states}
In order to calculate some matrix elements such as
$\langle D^*|J|B\rangle$ on the lattice, we have to prepare the
external states, such as $B$ meson state $|B\rangle$, for instance.
The operator to create such a state is not perfect, and a significant
amount of excited states with the same quantum number are created.
The common practice is then to wait for some (Euclidean) time until
the excited states die exponentially.
Or, one can fit the dependence on the time separation including the
effect of excited states and extract only the ground state.

\begin{figure}[tbp]
  \centering
  \includegraphics[width=8cm]{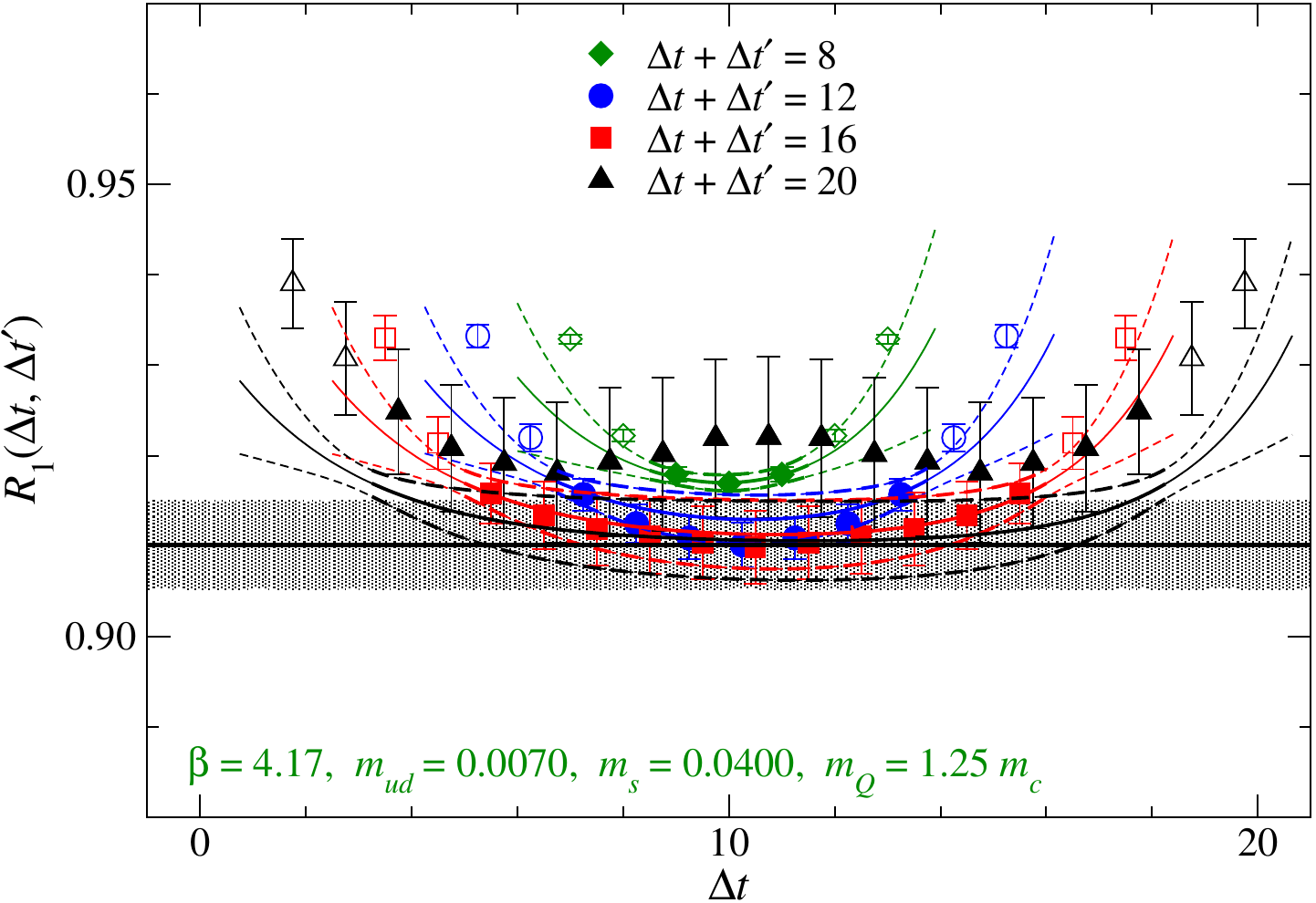}
  \caption{Extraction of the ground state from the three-point
    function to calculate the $B\to D^*$ form factor $h_{A1}(1)$.
    Data with various source-sink separation $\Delta t+\Delta t'$ are
    shown. 
    The plot is from JLQCD \cite{Aoki:2023qpa}.}
  \label{fig:R1}
\end{figure}

Fig.~\ref{fig:R1} shows an example of such attempt for three-point
functions to extract the $B\to D^*$ form factor $h_{A1}(1)$.
The time separation $\Delta t$ ($\Delta t'$) between the inserted
current and the $B$ meson ($D^*$ meson) interpolating field is varied
to find a plateau, the signal of the ground-state saturation.
The data themselves do not show convincing plateau, but by combining
them with various $\Delta t$ and $\Delta t'$ one can perform a global
fit assuming that the data are well saturated by the ground and
first-excited states. 
The different energy states are typically separated by the QCD scale
$\Lambda_{\mathrm{QCD}}$ or the pion mass $m_\pi$, so that one needs
the $\Delta t^{(\prime)}$ of about $1/\Lambda_{\mathrm{QCD}}$ or even
$1/m_\pi$, which can be prohibitively large ($\sim$ 10 in the unit
shown in Fig.~\ref{fig:R1}) given the exponentially degrading signal
for large $\Delta t$.

\begin{figure}[tbp]
  \centering
  \includegraphics[width=8cm]{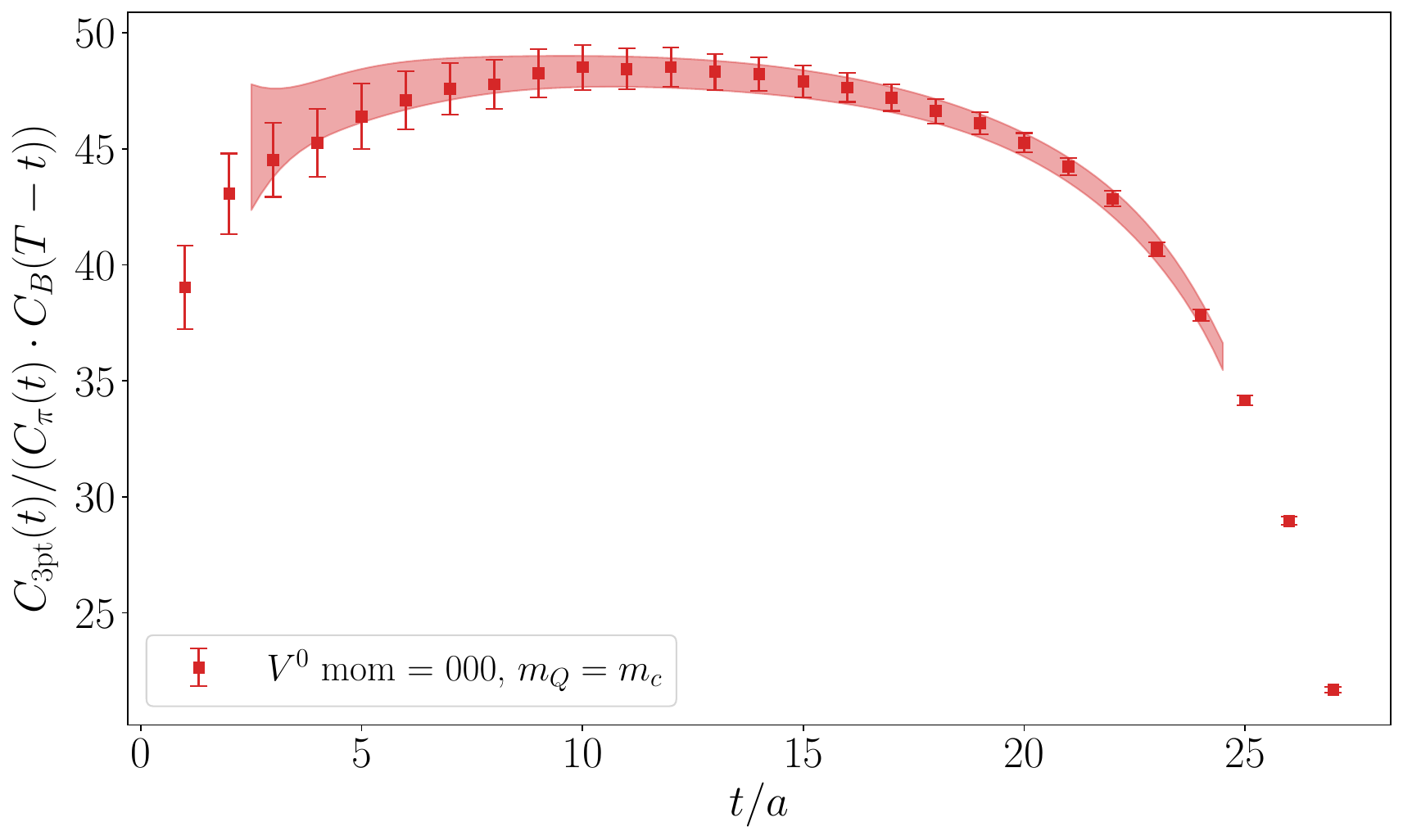}
  \caption{Extraction of the ground state from the three-point
    function to calculate the $B\to\pi$ form factor.
    The plot is from JLQCD \cite{Colquhoun:2022atw}.}
  \label{fig:R_B2pi}
\end{figure}

Another noticeable example is shown in Fig.~\ref{fig:R_B2pi}, which is
from the JLQCD calculation of the $B\to\pi$ form factor
\cite{Colquhoun:2022atw}. 
In the region of small time separation between the current of $B$
meson interpolation operator (large $t$ in the plot), we find a strong
deviation from a plateau.
A very similar deviation is also seen in the work of
Fermilab/MILC \cite{FermilabLattice:2015mwy} and
RBC/UKQCD \cite{Flynn:2015mha}.

Theoretical explanation of such significant excited-state
contamination has recently been attempted by
B\"ar, Broll, Sommer \cite{Bar:2023sef}
using heavy-meson chiral effective theory.
The contamination due to the $B\pi$ states is suggested to be the
source of the problem.
Such knowledge may be useful to suppress systematic errors in the
future lattice analyses.

\section{Next challenge III: $B\to D^{**}$}
Excited states are sometimes an important signal, not just a
contamination to eliminate.
An interesting example is the excited states of $D$ mesons appearing
in the $B$ meson semi-leptonic decays.
There are four $P$-wave states,
$D_0^*$, $D_1^*$, $D_1$ and $D_2^*$, collectively called $D^{**}$, and
there is a problem known as the ``1/2 vs 3/2 puzzle''
\cite{Bigi:2007qp}, which indicates that the experimentally observed
decay rates are not consistent with the theory prediction based on the
heavy quark effective theory. 
They also have a practical relevance to the $|V_{cb}|$ determination
as a potential background for the ground state
$B\to D^{(*)}\ell\bar\nu$ signals.

On the lattice, the study of the form factors involving excited states
is very limited, probably because the statistical signal for the
excited states is so poor to identify a 
plateau on the lattice.
A pioneering study by ETMC \cite{Atoui:2013ksa} suggests that it is
indeed the case.
More recently, another method to approach the problem was proposed
\cite{Bailas:2019diq,Hashimoto:2021hqu}, which utilizes the four-point
correlation functions calculated for the study of inclusive decays
\cite{Gambino:2020crt}.
Since the Compton amplitudes derived from the four-point functions
contain the amplitudes of all possible final states, the signal for
the $D^{**}$ mesons can be extracted, in principle.

\begin{figure}[tbp]
  \centering
  \includegraphics[width=8cm]{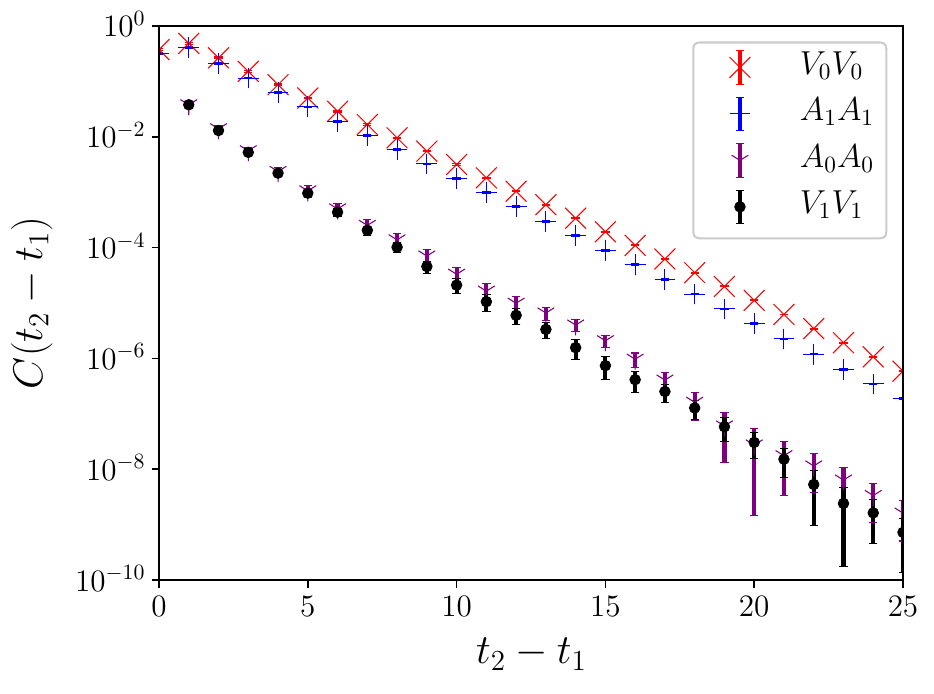}
  \caption{
    Compton amplitude $\langle B|J(t)J^\dagger(0)|B\rangle$
    corresponding to the decay rate of $B\to X_c$.
    The plot shows various channels (vector $V_\mu$ and axial-vector
    $A_\mu$; temporal ($\mu$ = 0) and spatial ($\mu$ = 1) components
    of the $b\to c$ currents) as a function of 
    the time separation between the $b\to c$ ($J^\dagger$) and $c\to
    b$ ($J$) current insertions. 
    The plot is from JLQCD \cite{Hashimoto:2021hqu}.}
  \label{fig:C4}
\end{figure}

An attempt is shown in Fig.~\ref{fig:C4}.
The Compton amplitude $\langle B|J(t)J^\dagger(0)|B\rangle$ is
calculated on the lattice from a corresponding four-point function.
All possible states are created between the two flavor-changing
currents $J^\dagger(0)$ and $J(t)$ depending on their quantum number.
For instance, in the zero-recoil limit $V_0$ and $A_k$ ($k$ stands for
a spatial direction) produce $S$-wave
pseudo-scalar $D$ and vector $D^*$ mesons, respectively.
With parity opposite currents $A_0$ and $V_k$, only $P$-wave states
may appear so that their amplitude is much smaller, yet their signal
is clearly visible.
Comparison of the zero-recoil form factor for $B\to D^{**}$ with the
heavy-quark effective theory expectation
\cite{Leibovich:1997em,Bernlochner:2016bci} has been attempted
\cite{Bailas:2019diq}.

\section{Next challenge IV: $B\to Kc\bar{c} \to K\ell^+\ell^-$}
In the search for new physics, the $B\to K^{(*)}\ell^+\ell^-$ mode
is a promising candidate, as it goes through a flavor-changing neutral
current (FCNC) and is highly suppressed in the Standard Model (SM).
Indeed, there is a well-known tension with the SM for an
angular observable, though the uncertainty in its SM prediction has
not been obtained from theoretically solid methods like lattice QCD.
Less significant but still suggestive tension is also observed for its
differential decay rate; the lattice calculation is used for its 
SM estimate \cite{Parrott:2022zte}.

For this decay mode, there is a potentially important source of
systematics originating from the charm loop, {\it i.e.}
$b\to sc\bar{c} \to s\ell^+\ell^-$.
The relevant four-fermion operator of the form $\bar{b}c\bar{c}s$ is
not suppressed and even highly enhanced by charmonium resonances in
the intermediate state.
To avoid the huge effect from charmonium ($J/\psi$ and $\psi'$), the
experimental analyses veto the corresponding mass region, but their
effects in the other nearby kinematical regions is not well understood.

The lattice calculation of such process is possible in principle.
But, one needs to treat two-body intermediate states like
$K^{(*)} \psi$ or $D D_s$, and the problem appears due to possible
intermediate states with an energy lower than the final state
$K^{(*)}\ell^+\ell^-$, {\it i.e.} the well-known Maiani-Testa no-go
theorem \cite{Maiani:1990ca} applies.
Therefore, it is an example of a broader challenge to overcome the
Maiani-Testa situation; some attempt has been proposed
\cite{Bruno:2020kyl}.

\section{No conclusion, yet}
As more precise experimental data become available from LHCb and
Belle II, pursuing better precision is a natural direction of lattice
QCD calculations.
Systematic errors need to be seriously investigated, and it appeared
that the heavy-quark extrapolation and excited-state contamination can
be a non-trivial issue.
They are in principle straight-forward to eliminate by simulating 
sufficiently fine lattices and long Euclidean time separations, but
the currently available lattice data (and those in the near future)
may be affected significantly.
The use of available theoretical ideas and constraints, such as those
from heavy-quark effective theory, would be very effective.
The data-driven approach would not always be bias free.

Next challenges in semi-leptonic $B$ decays include quantities beyond
the standard form factor calculations.
Decays to excited-state $D$ mesons poses an interesting
theoretical problem, and at the same time it would be important for the
experimental analysis.
Not covered in this talk is the inclusive decay rate, which opens a
broad range of new applications.

\vspace*{5mm}
I thank the present and past members of the JLQCD collaboration,
Takashi Kaneko in particular.
A lot of materials in this presentation emerged from the discussions
with them.
This work is partly supported by MEXT as ``Program for Promoting
Researches on the Supercomputer Fugaku'' (JPMXP1020200105)
and by JSPS KAKENHI, Grant-Number 22H00138.

\end{document}